\documentclass[fleqn,usenatbib]{mnras}

\usepackage{newtxtext,newtxmath}

\usepackage[T1]{fontenc}
\usepackage{ae,aecompl}

\usepackage{graphicx}	
\usepackage{amsmath}	
\usepackage{amssymb}	

\title[First evidence of diffuse ultra-steep-spectrum radio emission surrounding the cool core of a cluster]{First evidence of diffuse ultra-steep-spectrum radio emission surrounding the cool core of a cluster}

\author[F. Savini et al.] {F. Savini$^{1}$\thanks{E-mail: federica.savini@hs.uni-hamburg.de},
A. Bonafede$^{1,2}$, 
M. Br{\"u}ggen$^{1}$, 
R. van Weeren$^{3}$,
G. Brunetti$^{2}$,
H. Intema$^{3}$,
\newauthor 
A. Botteon$^{2,4}$
T. Shimwell$^{3,5}$,
A. Wilber$^{1}$, 
D. Rafferty$^{1}$,
S. Giacintucci$^{6}$,
R. Cassano$^{2}$,
\newauthor
V. Cuciti$^{2}$,
F. de Gasperin$^{4}$,
H. R{\"o}ttgering$^{3}$,
M. Hoeft$^{7}$,
G. White$^{8,9}$
\\
$^1$ Hamburger Sternwarte, Universit\"at Hamburg, Gojenbergsweg 112, 21029, Hamburg, Germany. \\
$^2$ INAF IRA, via P. Gobetti 101, 40129 Bologna, Italy.\\
$^3$ Leiden Observatory, Leiden University, PO Box 9513, 2300 RA Leiden, The Netherlands.\\
$^4$ Dipartimento di Fisica e Astronomia, Universita' di Bologna, via P. Gobetti 93/2, 40129 Bologna, Italy.\\
$^5$ ASTRON, the Netherlands Institute for Radio Astronomy, Postbus 2, 7990 AA, Dwingeloo, The Netherlands.\\
$^6$ Naval Research Laboratory, 4555 Overlook Avenue SW, Code 7213, Washington, DC 20375, USA.\\
$^7$ Th{\"u}ringer Landessternwarte, Sternwarte 5, 07778 Tautenburg, Germany.\\
$^8$ Department of Physics \& Astronomy, The Open University, Walton Hall, Milton Keynes MK7 6AA, UK.\\
$^9$ Space Science \& Technology Department, CCLRC Rutherford Appleton Laboratory, Chilton, Didcot, Oxfordshire OX11 0QX, UK.}

\date{Accepted XXX. Received YYY; in original form ZZZ}

\pubyear{2018}

\begin{document}
\label{firstpage}
\pagerange{\pageref{firstpage}--\pageref{lastpage}}
\maketitle

\begin{abstract}
Diffuse synchrotron radio emission from cosmic-ray electrons is observed at the center of a number of galaxy clusters. These sources can be classified either as giant radio halos, which occur in merging clusters, or as mini halos, which are found only in cool-core clusters. 
In this paper, we present the first discovery of a cool-core cluster with an associated mini halo that also shows ultra-steep-spectrum emission extending well beyond the core that resembles radio halo emission. The large-scale component is discovered thanks to LOFAR observations at 144 MHz. We also analyse GMRT observations at 610 MHz to characterise the spectrum of the radio emission. An X-ray analysis reveals that the cluster is slightly disturbed, and we suggest that the steep-spectrum radio emission outside the core could be produced by a minor merger that powers electron re-acceleration without disrupting the cool core. This discovery suggests that, under particular circumstances, both a mini and giant halo could co-exist in a single cluster, opening new perspectives for particle acceleration mechanisms in galaxy clusters. 
\end{abstract}




\section{Introduction}
An increasing number of diffuse radio sources associated with galaxy clusters are being detected with the advent of new facilities and techniques at low radio frequencies. Not only has the number of sources increased, but the quality of imaging in terms of resolution and noise has also improved, revealing various source morphologies and properties that might require a broadening of the taxonomy of radio emission in galaxy clusters (e.g. \citealp{Dega2017}). Diffuse emission in the form of giant radio halos or mini halos is found in the central regions of some galaxy clusters. These sources have low surface brightnesses and steep radio spectra\footnote{The radio spectrum follows a power law $S(\nu) \propto \nu^{\alpha}$, where $S$ is the flux density, and $\nu$ the observing frequency. Steep spectrum radio sources  spectra have spectral indices $\alpha < -1$.} that make them much brighter at lower frequencies. The Low Frequency Array (LOFAR; \citealp{VH2013}) operating between 30 and 240 MHz can discover steep-spectrum sources that are impossible to detect with other radio telescopes.\\

Giant radio halos have typical sizes of 1 to 2~Mpc, and are predominantly found in massive, merging clusters (e.g. \citealp{Buote2001}, \citealp{Cassa2010}, \citealp{Cuci2015}), suggesting that merger-driven turbulence re-accelerates primary or secondary electrons in the ICM (\citealp{Bru2007}, \citealp{Bru2011}, \citealp{Pin2017}). Mini halos have typical sizes of a few hundred kpc and a higher emissivity than giant halos (\citealp{Cassano2008}, \citealp{Murgia2009}). Mini halos are found exclusively in cool-core clusters, and are confined within the inner regions of the cluster (\citealp{Govo2009}, \citealp{Kale2015}, \citealp{Giaci2017}). 
Cool-core clusters display a peaked X-ray surface brightness and a significant drop in temperature ($< 10^7 - 10^8$ K) at the centre. Signatures of minor mergers have been detected in some cool-core clusters which host mini halos \citep{MG2008}. When the low-entropy central gas at the bottom of the dark matter potential well is perturbed by a minor merger and meets the higher-entropy ICM, a discontinuity in the X-ray emissivity, called a cold front, is formed. The gas then falls back into the dark matter potential well and ``sloshes", possibly generating the turbulence that re-accelerates weakly relativistic electrons within the core \citep{ZuHone2013}. Particle acceleration by turbulence is an inefficient mechanism and according to theoretical models only major mergers between massive clusters can dissipate enough energy to power radio emission on Mpc scales up to GHz frequencies. 
So far, it is unknown what happens when a cool-core cluster hosting a mini halo undergoes a minor merger that does not disrupt the core. This scenario is particularly interesting when observed at low radio frequencies, as it may provide new insights on the connections between mini halos, giant radio halos, and the cluster dynamics.\\

 In this paper we report on the results of a LOFAR radio observation of the galaxy cluster PSZ139139.61+24.20. We assume a flat, $\Lambda$CDM cosmology with matter density $\Omega_M = 0.3$ and Hubble constant $H_0 = 67.8$ km s$^{-1}$ Mpc$^{-1}$ \citep{Planck2016}. The angular to physical scale conversion at z = 0.267 is 4.137 kpc/$''$. All our images are in the J2000 coordinate system.

\begin{table}
 \centering
 \caption{Properties of the galaxy cluster PSZ1G139.61+24.20 \citep{Giaci2017}. (1),(2),(3),(4): Target coordinates;  (5): Redshift; (6): Global temperature computed within the radius enclosing a mean density of 2500 times the critical density at the cluster redshift. Note that the central region with a radius of 70 kpc was excised; (7): Core entropy; (8): Mass within the radius enclosing a mean density of 500 times the critical density \citealp{Planck2014}; (9): Radius that encloses a mean overdensity of 500 with respect to the
critical density at the cluster redshift.} 
  \begin{tabular}{c c}
  \hline
 1: RA \scriptsize{(h:m:s)}	   &  06:22:13.9  \\
 2: DEC \scriptsize{($^\circ$:$'$:$''$)} &  +74:41:39.0\\
 3: $l$ \scriptsize{($^\circ$)}& 95.57 \\
 4: $b$ \scriptsize{($^\circ$)}&  74.69 \\
 5: $z$  &  $0.267$\\
 6: kT \scriptsize{(keV)} &  $7.5 \pm 0.4 $\\
 7: K$_0$ \scriptsize{(keV \,cm$^2$)} &  $< 20$\\
 8: M$_{500}$ \scriptsize{($\times 10^{14}$ M$_{\odot}$)} & $7.1 \pm 0.6 $  \\
 9: R$_{500}$ \scriptsize{(Mpc)} & 1.24 \\ 
\hline
\end{tabular}
\label{info}
\end{table}

\subsection{The cluster}

PSZ1G139.61+24 (z = 0.267, RA = 06:22:13.9, DEC = +74:41:39.0; hereafter PSZ139) has been classified as a galaxy cluster through detection of the Sunyaev-Zel'dovoch effect with the Planck satellite \citep{Planck2014}. Using GMRT observations at 610 MHz, \citet{Giaci2017} report the detection of a tentative mini halo with an overall source size of $\sim 100$ kpc located at the cluster centre. They also present density and temperature profiles derived from a Chandra X-ray observation of PSZ139. Details can be found in Tab.~\ref{info}. The cluster core is characterized by low values of temperature, and the temperature profile inverts and starts decreasing approaching the cluster centre within a radius of $\sim 100$ kpc (see Fig.2 in \citet{Giaci2017}). The specific entropy \citep{Cava2009} at the cluster centre, $K_0$, is used to distinguish between cool-core or non cool-core clusters \citep{Giaci2017}: clusters with low central entropies ($K_0 < 30 - 50$ keV cm$^2$) are expected to host a cool core. The specific entropy of PSZ139 is $K_0 < 20$ keV cm$^2$ \citep{Giaci2017}, indicating that this cluster has a cool core.

\section{Data reduction}
\label{sec:radio}

\begin{table*}
 \centering
  \caption{Col. 1: Telescope/Survey; Col. 2: Central frequency; Col. 3: Minimum baseline; Col. 4: Largest angular scale; Col. 5: Resolution; Col. 6: rms noise level; Col. 7: Parameters used for LOFAR and GMRT imaging, such as gaussian taper (T) and weighting scheme; when Briggs weighting scheme is used, the robust value is specified \citep{Briggs}.}
 \begin{tabular}{c c c c c c c}
  \hline
1: Telescope & 2: Freq. &  3: $B_{min}$ & 4: $LAS$ & 5: Res. & 6: rms & 7: Imaging\\
 & \scriptsize{(MHz)} &  \scriptsize{($\lambda$)} & \scriptsize{($'$)} & & \scriptsize{($\mu$Jy/beam)} & \\
 LOFAR & 144 & 80  & 43 & 5$'' \times$ 5$''$  & 140 & Briggs -0.25\\
  & & 80 & 43 & 11$'' \times$ 8$''$  & 150 & Briggs -0.25\\
  & & 80  & 43 & 35$'' \times$ 35$''$  & 500 & Briggs 0, 20$''$ T\\
 &  & 200 & 17 & $18'' \times 18''$   & 240 & uniform, 20$''$ T\\
  GMRT & 610 & 150  & 23 & $8.0'' \times 4.7''$ & 27 & Briggs -0.1\\
   &  & 150  & 23 &32$'' \times 32''$ & 180  & Briggs 0, 20$''$ T\\
   &  & 200 &  17 &$20'' \times 20''$ & 130  & uniform, 20$''$ T\\
\hline
\end{tabular}
\label{obs}
\end{table*}

\subsection{LOFAR radio observation}

The cluster PSZ139 was observed as part of the LOFAR Two-Metre Sky Survey (LoTSS; \citealp{Shim2017}) at High Band Antenna (HBA) frequencies (120 - 168 MHz). The observation was carried out on July 27, 2017 (ID LC8\_022) with a total on-source time of 8 h preceded and followed by a flux calibrator (3C295) observation of 10 min. The calibration and imaging procedure is based on the facet calibration scheme presented in \citet{vanWeeren2016}. A complete outline of the procedure can be found in \citet{Savini2018}; here we will only briefly summarize the main steps:
\begin{itemize}
\item Preliminary pre-processing was performed by the Radio Observatory (ASTRON) and has been applied to the data;
\item Initial calibration was performed using the standard LOFAR direction independent calibration pipeline\footnote{https://github.com/lofar-astron/prefactor}; 
\item Flagging was performed after inspecting the dataset; bad data were found and flagged for a total of 30 min;
\item To refine the calibration, a pipelined version of the direction dependent facet calibration procedure was used\footnote{https://github.com/lofar-astron/factor}.
\end{itemize}

In facet calibration the field of view is divided up into a discrete number of directions (facets) that are separately calibrated through the selection of a calibrator (with a minimum flux of 0.5 Jy) for each facet. The coordinates of PSZ139 and a 15$'$ radius around it were also specified to include the source in one single facet. 
We processed 13 facets, i.e. the brightest sources in the field and those bordering the facet containing PSZ139, which was then processed at last, so that it could benefit from the calibration of the preceding facets. All the images were corrected for the station primary beam.\\

Due to inaccuracies in the LOFAR beam model the images can require rescaling (e.g. \citealp{Har2016}). In line with other LOFAR studies \citep{vanW2014}, we have cross-checked the 144 MHz LOFAR flux scale against the 150 MHz TIFR GMRT Sky Survey (TGSS; \citealp{Int2017}) using 50 compact sources. We found and applied a scaling factor of $0.75$ with a scatter that we take into account by assigning a $15\%$ uncertainty in our flux scale. The corrected integrated surface brightness measurements are reported in Table \ref{fluxes}.\\

Radio imaging was performed using the Common Astronomy Software Applications (CASA, version 4.5.2; \citealp{Mc2007}) tools with different parameters to obtain different resolutions and increase the sensitivity to diffuse emission. The imaging details are summarized in Table \ref{obs}.\\

\subsection{GMRT radio observation}

GMRT observations at 610~MHz were collected during two distinct observations on October 25, 2014, and September 4, 2015, under project codes 27\_025 and 28\_077, respectively. Visibilities were recorded in two polarizations (RR and LL) over a bandwidth of 33.3~MHz. The total combined on-source time was 9.4 h. The GMRT data were pre-calibrated, combined and processed using the SPAM pipeline (see \citealp{Int2017} for details). The primary calibrators used for flux and bandpass calibrations were 3C147 and 3C48, respectively. We adopted the same flux standard as for LOFAR \citep{SH2012}. A $T_{\rm sys}$ correction of $0.92$ was derived using the all-sky map at 408 MHz by \citet{Haslam1995}, and applied. Both observations had 4 of the 30 antennas not working properly. Removing them during pre-calibration resulted in a 25\% data loss. Furthermore, the pipeline removed another 15\% of the data due to RFI and various telescope issues. The pipeline output visibilities were imported into CASA for final imaging, using the multi-scale option of the \texttt{clean} task. Our highest-fidelity images reach a sensitivity of $27 \, \mu$Jy/beam with a $8.0'' \times 4.7''$ beam. We adopted a 10\% scale error on all flux density measurements \citep{Chandra2004}.

\begin{figure}
\includegraphics[width=0.5\textwidth]{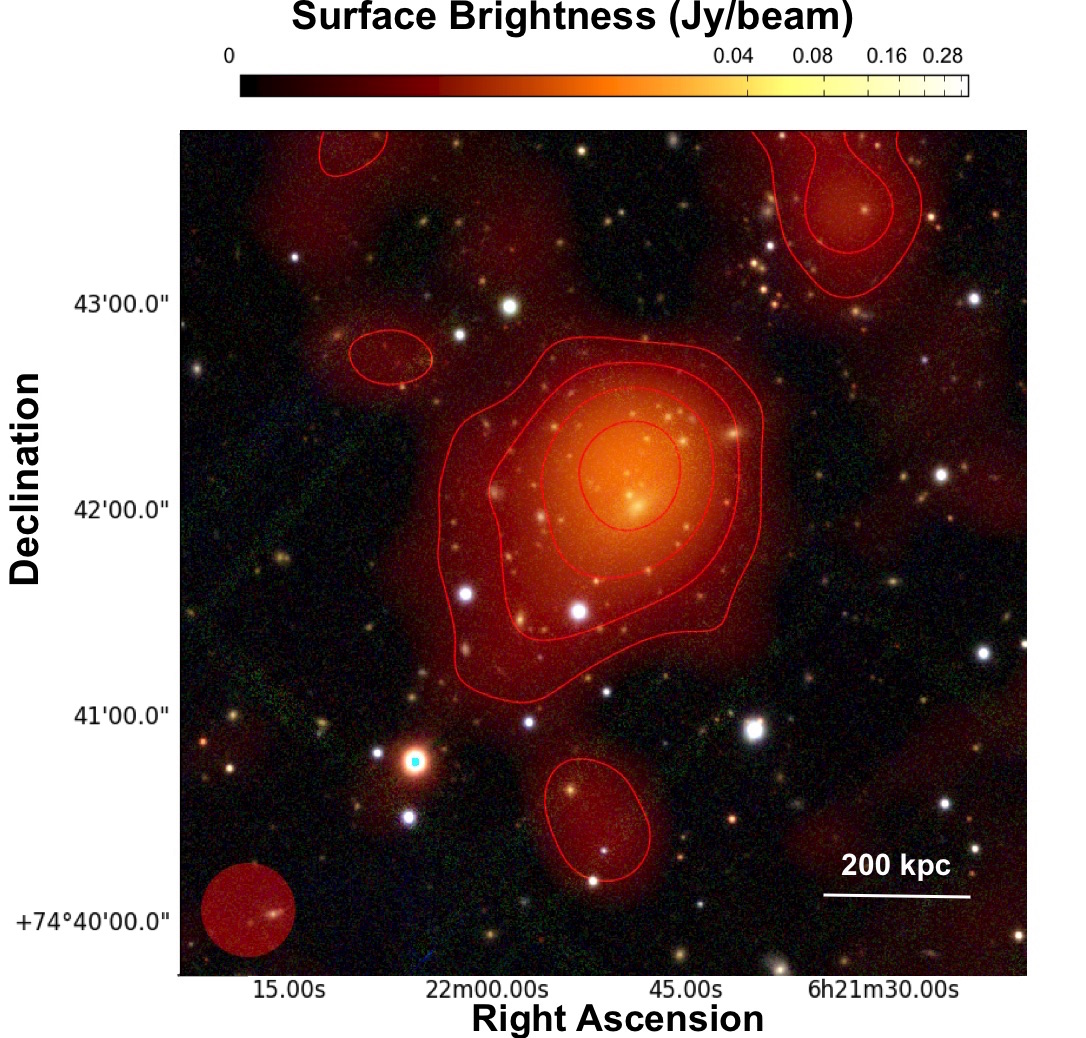}
 \caption{ Optical Pan-STARRS g,r,i mosaic image with the 144 MHz LOFAR smoothed image overlaid. The contour levels are at $(-1, 1, 2, 4, 8)\, \times \, 3\sigma$ where $\sigma$ = 500 $\mu$Jy/beam. The beam is 35$'' \times$ 35$''$, and is shown at the bottom left of the image.}
 \label{LOFAR1}
\vspace{2cm}
\includegraphics[width=0.5\textwidth]{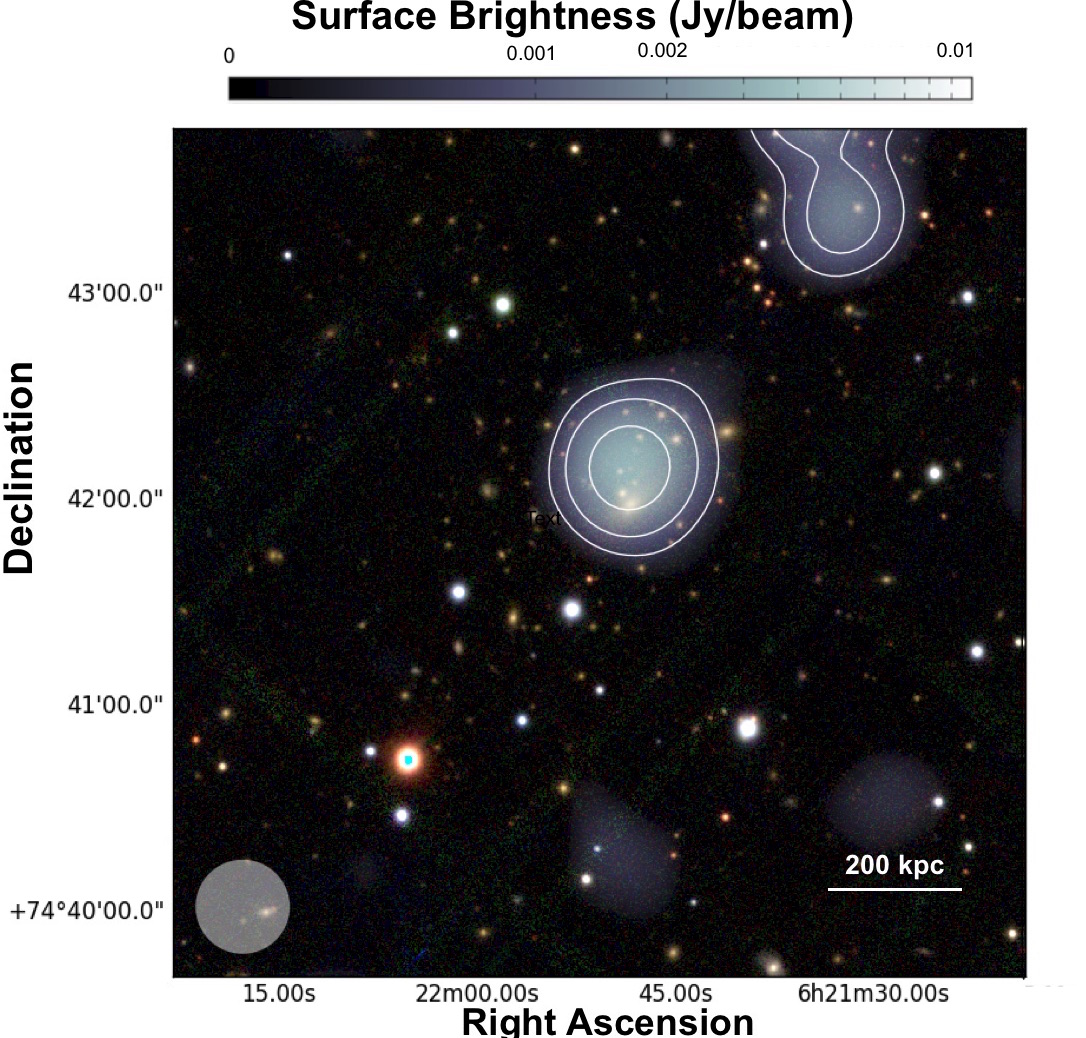}
   \caption{Optical Pan-STARRS g,r,i mosaic image with the 610 MHz GMRT smoothed image overlaid. The contour levels are at $(-1, 1, 2, 4)\, \times \, 3\sigma$ where $\sigma$ = 180 $\mu$Jy/beam. The beam is 35$'' \times$ 35$''$, and is shown at the bottom left of the image.}
\label{LOFAR2}
\end{figure}

\subsection{Chandra X-ray observation}

We reprocessed Chandra X-ray observations merging two Chandra ACIS-I observations of PSZ139 in \texttt{VFAINT} mode with a total exposure time of 28 ks (ObsID: 15139, 15297). Data were reprocessed with
CIAO v4.9 and Chandra CALDB v4.7.3 using the \texttt{level=1} event file, following the standard Chandra reduction threads. Soft proton flares were removed by inspecting the light curves extracted in the S2 chip using the \texttt{deflare} script. The resulting exposure time after this procedure is 23.1 ks. A single point spread function map at 1.5 keV was obtained combining the corresponding exposure maps for each ObsID, then point sources were detected with \texttt{wavdetect}, confirmed by eye and removed in the further analysis. Spectra were fitted in the 0.5-11.0 keV band with XSPEC v12.9.0o adopting an absorbed thermal model with metallicity fixed at 0.3 $Z_{\odot}$ for the ICM emission and with a fixed column density \citep{Kalberla2005} $N_{\rm H} = 8.1 \times 10^{20}$ cm$^{-2}$ accounting for the Galactic absorption in the direction of the cluster. The background was treated as follows: the astrophysical background was assumed to be composed of a Galactic component, modeled with a two temperature-plasma (with $kT_{1} = 0.14$ keV and $kT_2 = 0.25$ keV), and a cosmic X-ray background component, modeled with an absorbed power-law (with $\Gamma = 1.4$); the instrumental background was modeled following the analytical approach proposed in \citet{Bar2014}. The X-ray analysis follows the procedure described in \citet{Botteon2018}, to which we refer the reader for more details.

\section{Results}
\label{res}

\subsection{Radio analysis}
Using LOFAR observations at 144 MHz, we have discovered previously undetected cluster-scale diffuse emission in PSZ139, as visible in Fig. \ref{LOFAR1}. We have re-analysed two archival 610 MHz GMRT observations, and combined their visibilities to achieve better sensitivity and {\it uv}-coverage. In Fig. ~\ref{LOFAR1} and ~\ref{LOFAR2}, we present the low-resolution LOFAR and GMRT radio images. The images were smoothed to enhance the diffuse emission, and are overlaid onto Pan-STARRS g,r,i optical images \citep{Cha2016}. 
In Fig.  \ref{high}, we present the high-resolution radio contours in the core region of the cluster. This consists of two radio components: one at the cluster centre where also the X-ray peak is, and one towards the N. The central source is likely to be related to the Brightest Central Galaxy (BCG) that is visible in the optical image, whilst there is no obvious optical counterpart for the northern radio brightness source that might be a foreground radio galaxy. For our analysis, we have considered the X-ray centre as the cluster centre.\\
Although both GMRT and LOFAR detect emission in the inner 200 kpc, i.e. within the core, LOFAR reveals a more extended component which is not detected in the GMRT image.  To confirm the detection of radio diffuse emission from PSZ139, we re-imaged the GMRT residual visibilities (after subtracting the full-resolution model image) with a gaussian taper of 25$''$ while enhancing the contribution of the short baselines with a Briggs weighting scheme (robust=0). We detected emission above $3 \sigma$ on a scale marginally larger than the cluster core ($\sim 350$ kpc, having a beam size equivalent to 75 kpc). Since we do not detect emission that corresponds to the halo observed with LOFAR, and since instrumental differences of the two observations must be taken into account, we did not use this image to derive spectral index information, but proceeded with a more conservative approach, as explained in Section  \ref{spix}.\\

From the LOFAR image, we measure an overall source size of $D_{\rm radio} \sim$ 550 kpc. This value has been estimated as $D_{\rm radio} = \sqrt{D_{\rm min} \times D_{\rm max}}$,  where $D_{\rm min}$ and $D_{\rm max}$ are the minimum and maximum diameter of the 3$\sigma$ surface brightness isocontours. Since this value may depend on the sensitivity of the observation, we have also estimated the e-folding radius, $r_{\rm e}$, which is defined as the radius at which the brightness drops to $I_{\rm 0}/e$, where $I_{\rm 0}$ is the central brightness of the source. Following \citet{Murgia2009}, we have obtained the radio brightness average in concentric circular annuli centred on the X-ray centre with widths of 12$''$ ($\sim$ 50 kpc) that is 1/2 FWHM of the synthesized beam, and assumed a profile that follows the simple exponential law $I(r) = I_{\rm 0} \, e^{-r/r_e}$. We convolved the exponential profile with a gaussian with FWHM equal to the beam, obtaining the convolved profile that we used to fit the data points shown in Fig.  \ref{profile}. The error of each annulus is equal to $\sqrt{\sigma_{\rm flux}^2 + \sigma^2 \times N}$ where the first contribution is the calibration error on the surface brightness and the second one is the noise level of the radio image weighted by the number of beam in the annulus. The best-fit values are $r_e = 94 \pm 10$ kpc and $I_0$ = $ 48 \pm 2 \, \mu$Jy/arcsec$^2$. The fit shows a relatively compact ($r_{\rm e} \sim 94$ kpc) emission in the core region. Larger-scale emission, although faint, can be seen beyond the core region, especially extending towards the SE.

\begin{figure}
\includegraphics[width=0.45\textwidth]{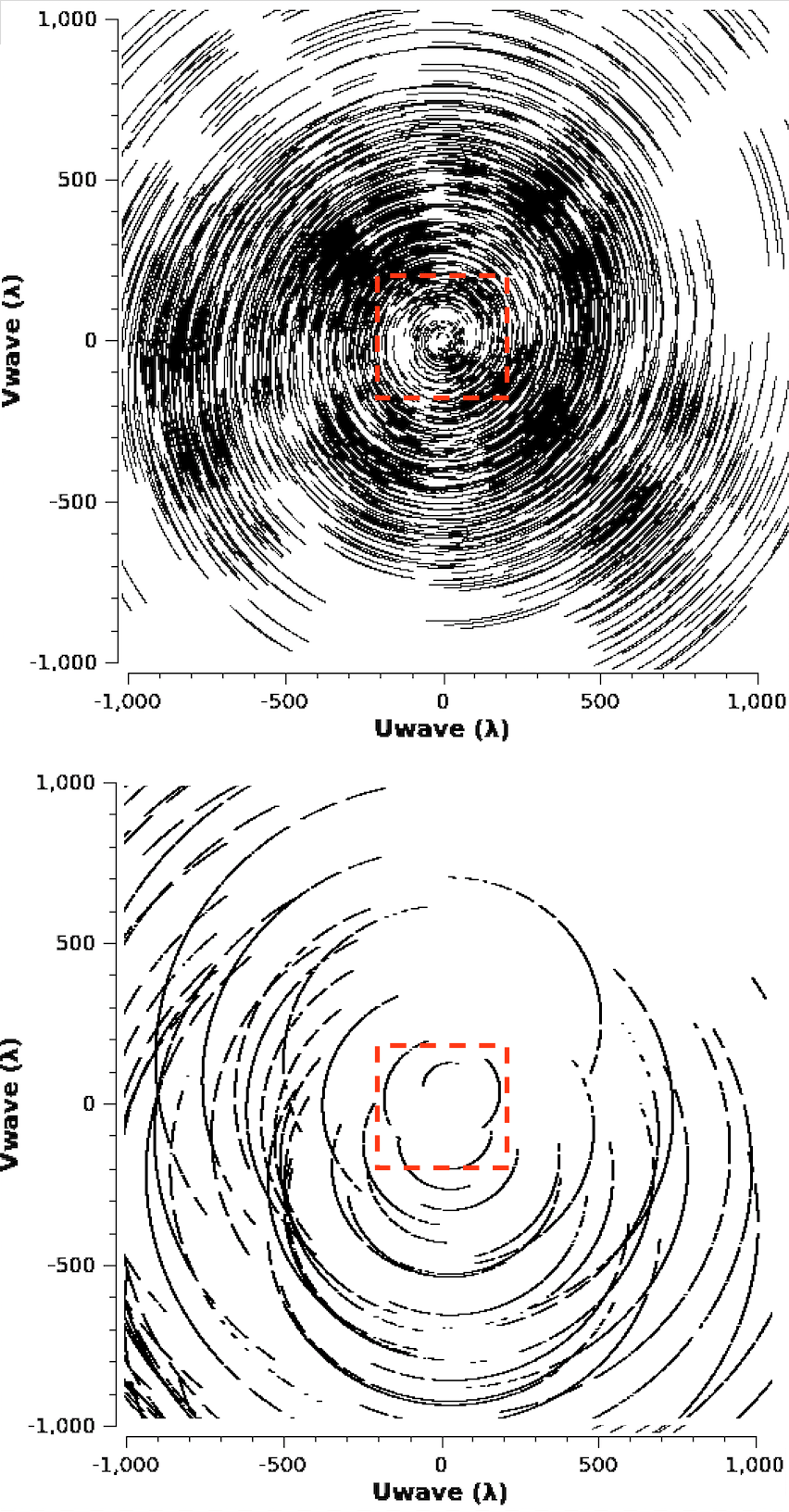}
  \caption{Plots of the inner \textit{uv}-plane coverage of the LOFAR observation (three time chunks are plotted; top panel) and GMRT observation (bottom panel). The red box indicates the region within 200$\lambda$ that was excluded in the imaging process for the spectral analysis. To minimise the difference between the two observations we have also used a uniform weighting parameter.}
\label{uv}
\end{figure}

 \begin{figure}
	 \includegraphics[width=0.48\textwidth]{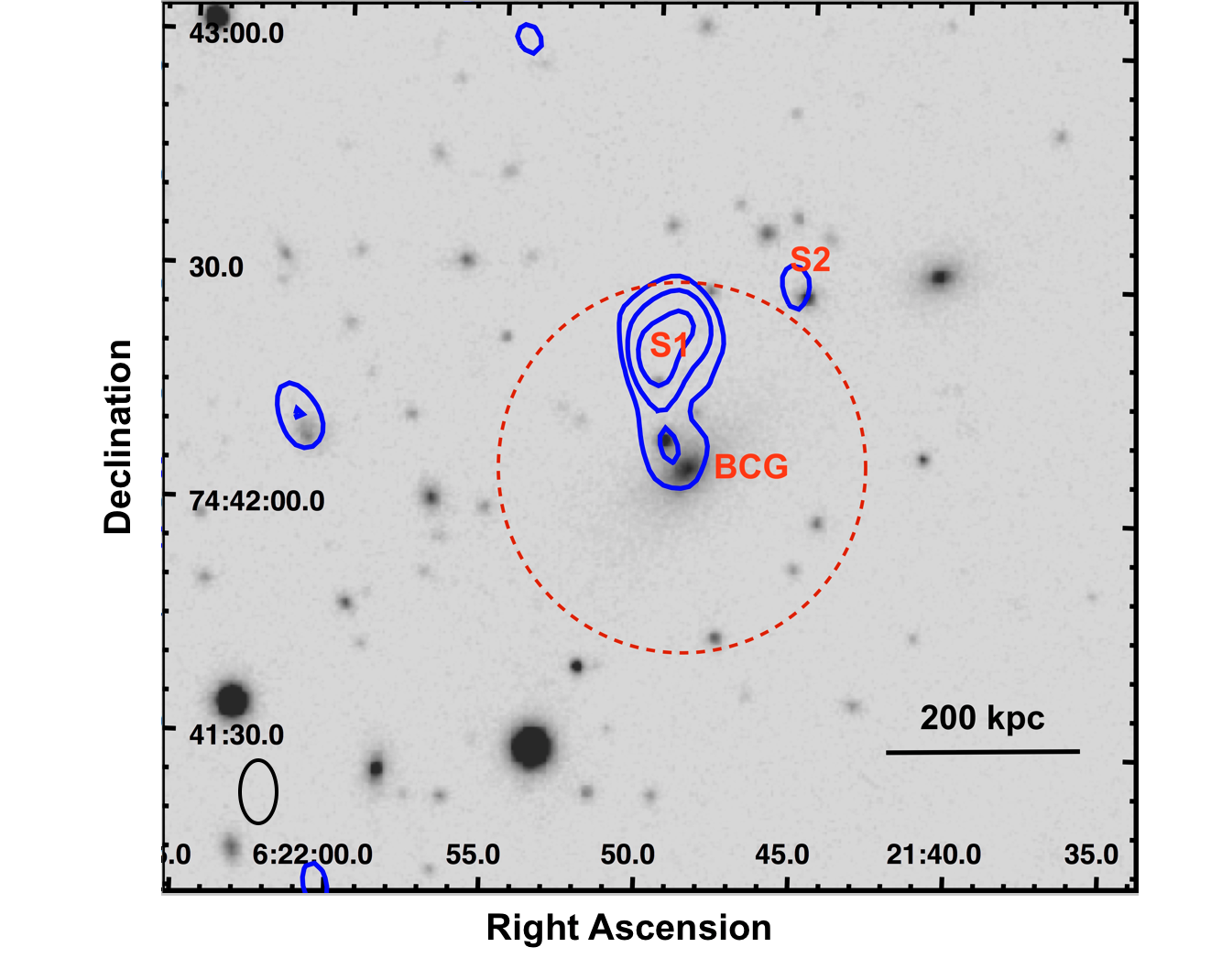}
     \includegraphics[width=0.47\textwidth]{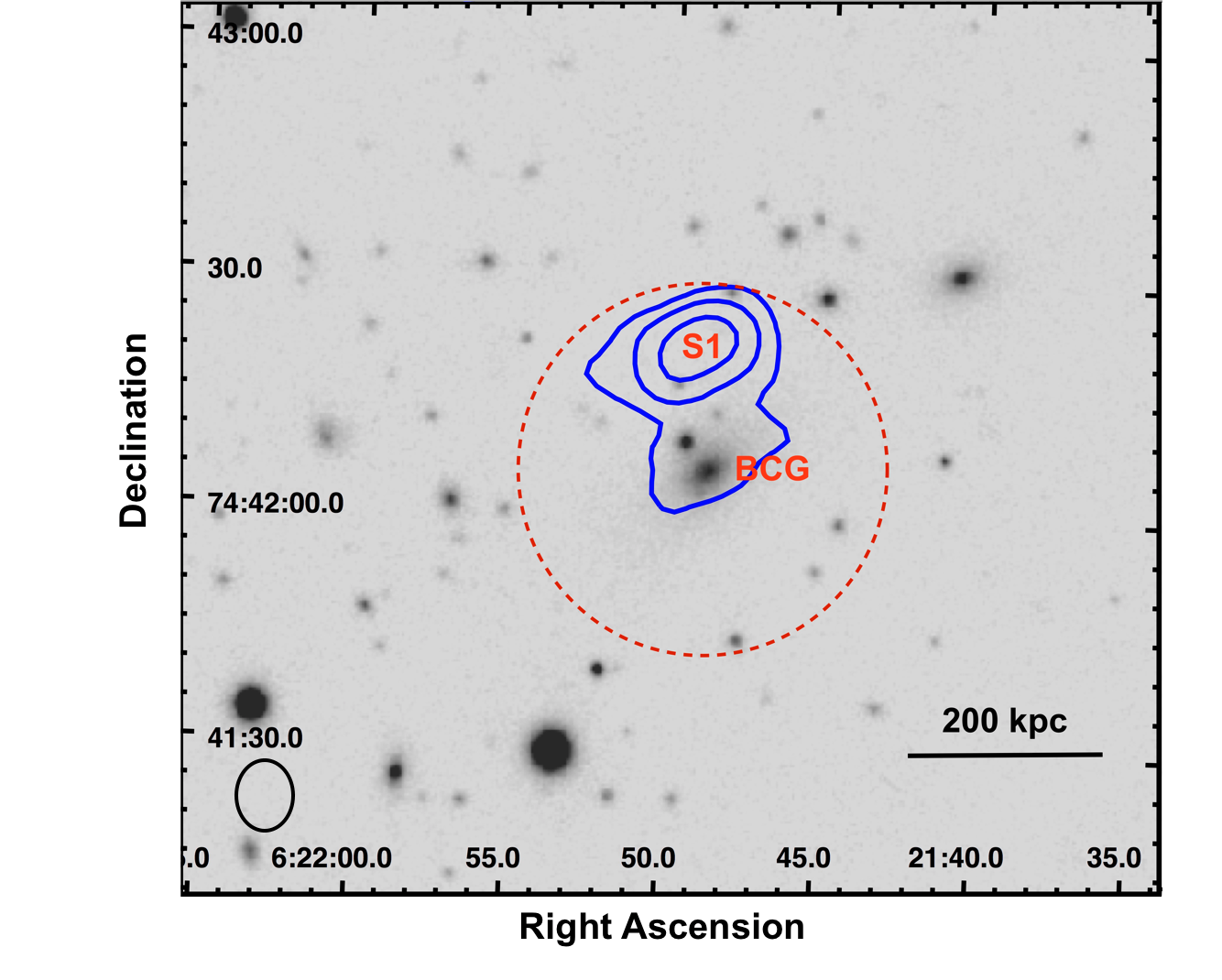}
    \caption{Optical Pan-STARRS image in gray scale with the high-resolution 610 MHz GMRT (top) and 144 MHz LOFAR (bottom) blue contours overlaid. The GMRT and LOFAR beam is $8.0'' \times 4.7''$ and $8.0'' \times 6.5''$ respectively, and the GMRT and LOFAR levels are $(2, 4, 8)\, \times \, 3\sigma$ where $\sigma$ = 27 $\mu$Jy/beam and $\sigma$ = 140 $\mu$Jy/beam respectively. The central radio components are indicated with labels: BCG is the brightest central galaxy that corresponds also to the X-ray centre; S1 is a source that has no obvious optical counterpart, and is likely to be a background galaxy; S2 is a faint source that is not detected in the LOFAR image. The dashed red region indicated the core region of the cluster, which has a size of $\sim 200$ kpc.}
    \label{high}
\end{figure}

\begin{table}
 \centering
  \caption{Integrated surface brightness and estimated radio power of the radio emission of PSZ139. The first two coloumns refer to values measured from the LOFAR and GMRT datasets. The last two coloumns refer to the value estimated for the surface brightness rescaled to 1.4 GHz, and the spectral index value used for rescaling. The core component is defined as the emission from the inner region with a size of $\sim 200$ kpc; the diffuse component corresponds to the emission on larger scales. The total value refers to the emission as a whole. }
\begin{tabular}{c c c c c}
  \hline
& LOFAR & GMRT & 1.4 GHz & $\alpha$\\
$S_{\rm core}$ (mJy) & 12 & 2.3 & - & \\
$P_{\rm core}$ (W/Hz)&  $3 \times 10^{24}$ & - &  $1.5 \times 10^{23}$ & -1.3 \\
$S_{\rm diff}$ (mJy) & 18 &-& & \\
$P_{\rm diff}$ (W/Hz) & $4 \times 10^{24}$ &-& $2.2 \times 10^{23}$ & -1.7 (UL)\\
$S_{\rm tot}$ (mJy) & 30 &-& & \\
$P_{\rm tot}$ (W/Hz) & $7 \times 10^{24}$ &- & $3.7 \times 10^{23}$ & -1.3\\
\hline
\end{tabular}
\label{fluxes}
\end{table}

\subsubsection{Spectral analysis} 
\label{spix}

We have re-imaged the LOFAR and GMRT datasets with a guassian taper of 20$''$, same pixel size, baseline range (200 - 40000 $\lambda$) and uniform weighting scheme to minimize the differences in the {\it uv}-coverage of the two interferometers.\\

Since only the central region of the cluster can be seen both at 610 MHz and 144 MHz, we have measured the value of the average spectral index of the inner $D_{\rm radio} \sim 200$ kpc. To ensure that no contamination from AGN or background sources was included in our estimate of the spectral index of the core region, we measured the integrated surface brightness of the compact sources detected in the high resolution image within the 6$\sigma$ GMRT and LOFAR contours, as shown in Fig.  \ref{high}; we then measured the integrated surface brightness of the entire core region (inner $\sim $200 kpc, indicated by a dashed red circle), and finally we subtracted the contribution of the compact sources\footnote{The best strategy would be subtracting the compact sources from the visibilities of the LOFAR and GMRT observations. However, in this case this procedure is uncertain, since these compact sources may have extended components (e.g. lobes) that are not easily separable from the surrounding emission.}. We obtained $\alpha_{144}^{610} = -1.3 \pm 0.1$. The error takes into account the flux calibration error. To confirm this value, we also calculated the spectral index, completely masking the sources at $6\sigma$, obtaining a consistent value.\\

To constrain the spectral properties of the diffuse emission (i.e. emission outside the inner $\sim 200$ kpc) that was detected in the LOFAR image only, we have also used the LOFAR and GMRT dataset re-imaged with uniform weighting and same {\it uv}-range mentioned above. We first considered the LOFAR mean surface brightness of the diffuse emission ($3.5$ mJy/beam) and the GMRT rms noise ($0.17$ mJy/beam), deriving $\alpha_{144}^{610} \leq -1.9$. The inner {\it uv}-coverage of the GMRT and LOFAR datasets are different (see Fig. \ref{uv}). Although the radio emission extends on a scale of 2$'$, that is well sampled by both observations, we have injected a mock radio halo in the GMRT visibilities. Using this procedure, we can image the diffuse emission and place an upper limit on the spectral index given the specific {\it uv}-coverage of that observation. 
The mock source was modeled with the exponential law and the parameters obtained from the best-fit of the radio surface brightness profile. The model was Fourier transformed into the visibilities of the GMRT dataset taking into account the w-projection effect, which is necessary due to the large field of view and low frequency. We added the mock sources to the original visibilities in a region close to the cluster but without bright sources and clear noise structures, such as negative holes, and then re-imaged the dataset with uniform weighting and measured the properties of the recovered simulated emission.
We created a set of mock sources assuming different spectral indices, i.e. with different integrated flux densities. We started with $\alpha_{144}^{610} = -1$ and then we lowered the value, until the recovered flux of the mock source could not be considered detected anymore, i.e. when the emission was $< 2\sigma$ and the extension $< 3r_e$. Using this procedure, we put an upper limit of $\alpha < -1.7$.\\

On the basis of the spectral information and surface brightness we derived in our radio analysis, we can distinguish two components of the radio emission:

\begin{itemize}
\item a flatter-spectrum ($\alpha_{144}^{610} = -1.3 \pm 0.1$), higher-brightness ($\sim$ 8 - 9 mJy/beam at 144 MHz) component within the cluster core with a size of $D_{\rm radio} \sim 200$ kpc, classified as mini halo by \citet{Giaci2017};

\item a steeper-spectrum ($\alpha_{144}^{610} < -1.7$), lower-brightness ($\sim$ 1 - 2 mJy/beam at 144 MHz) component visible on larger scales at low frequencies. 
\end{itemize}

\subsection{X-ray analysis}

\begin{figure}
	\includegraphics[width=0.45\textwidth]{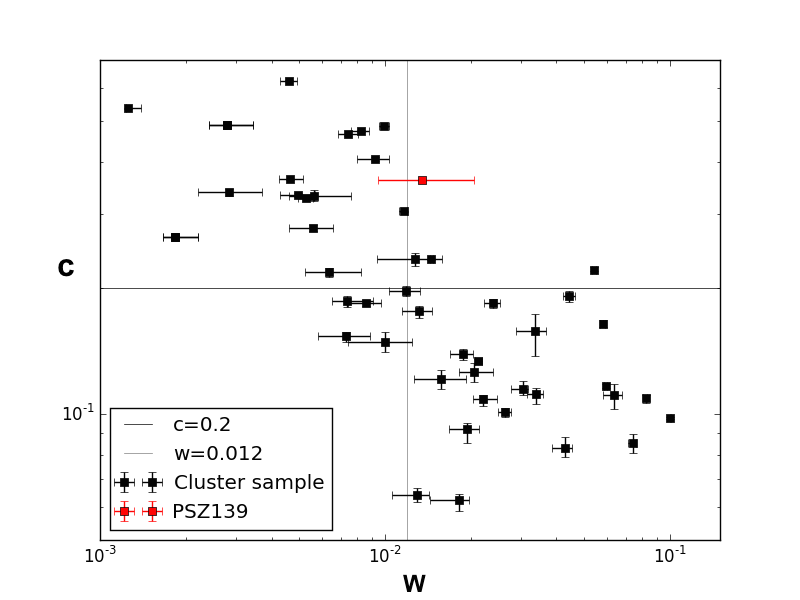}
    \caption{Diagram of the X-ray morphological indicators based on Chandra observations for the galaxy clusters, $w$ and $c$, of the mass-selected cluster sample in \citet{Cuci2015} and \citet{Cassa2016}. The red square represents PSZ139. Following \citet{Cassa2010}, we have adopted the values $w \le 0.012$ and $c \ge 0.2$ to separate merging from non-merging clusters. Merging clusters lie in the lower right region of the plot, whilst non-merging clusters in the upper left region.}
    \label{wc}
\end{figure}

Merger activity in clusters leaves a clear imprint on the X-ray brightness distribution, hence the morphology of the ICM provides a way to discriminate between merging and non-merging clusters. Here, we focus on two morphological indicators that can be derived from the X-ray surface brightness distribution \citep{Cuci2015}: the emission centroid shift $w$ and the concentration parameter $c$. The former is defined as the standard deviation of the projected separation between the peak and centroid of the X-ray surface brightness distribution when the aperture used to compute it decreases from a maximum radius of 500 kpc to smaller radii. The latter is defined as the ratio of the X-ray surface brightness within a radius of 100
kpc over X-ray surface brightness within a radius of 500 kpc. High values of $w$ indicate a dynamically disturbed system, whilst high values of $c$ indicate a peaked core, typical of non-merging systems. The position of PSZ139 in the $w - c$ diagram, based on Chandra X-ray observations, is reported in Fig. ~\ref{wc}, where we see that despite the high value of the concentration parameter ($c = 0.362 _{-0.004}^{+0.007}$), the value of the emission centroid shift ($w = 1.35_{-0.17}^{+0.18}  \times 10^{-2}$) is intermediate between merging and non-merging systems \citep{Cassa2016}. These values strengthen the argument that the cluster hosts a cool core (as also suggested by the low central entropy of the ICM) but is not fully virialised (relaxed).\\

 \begin{figure*}
  \includegraphics[width=0.44\textwidth]{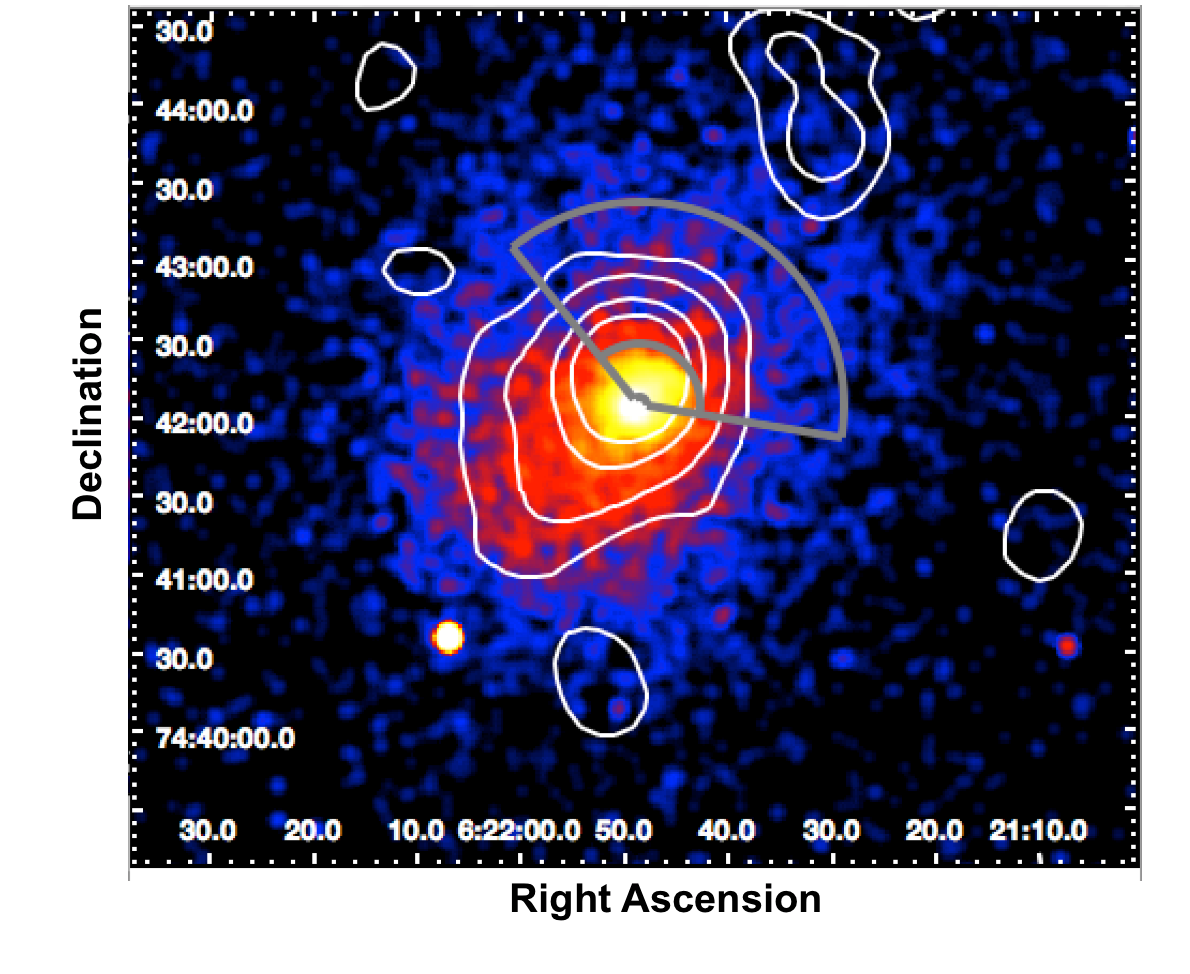}
        \includegraphics[width=0.45\textwidth]{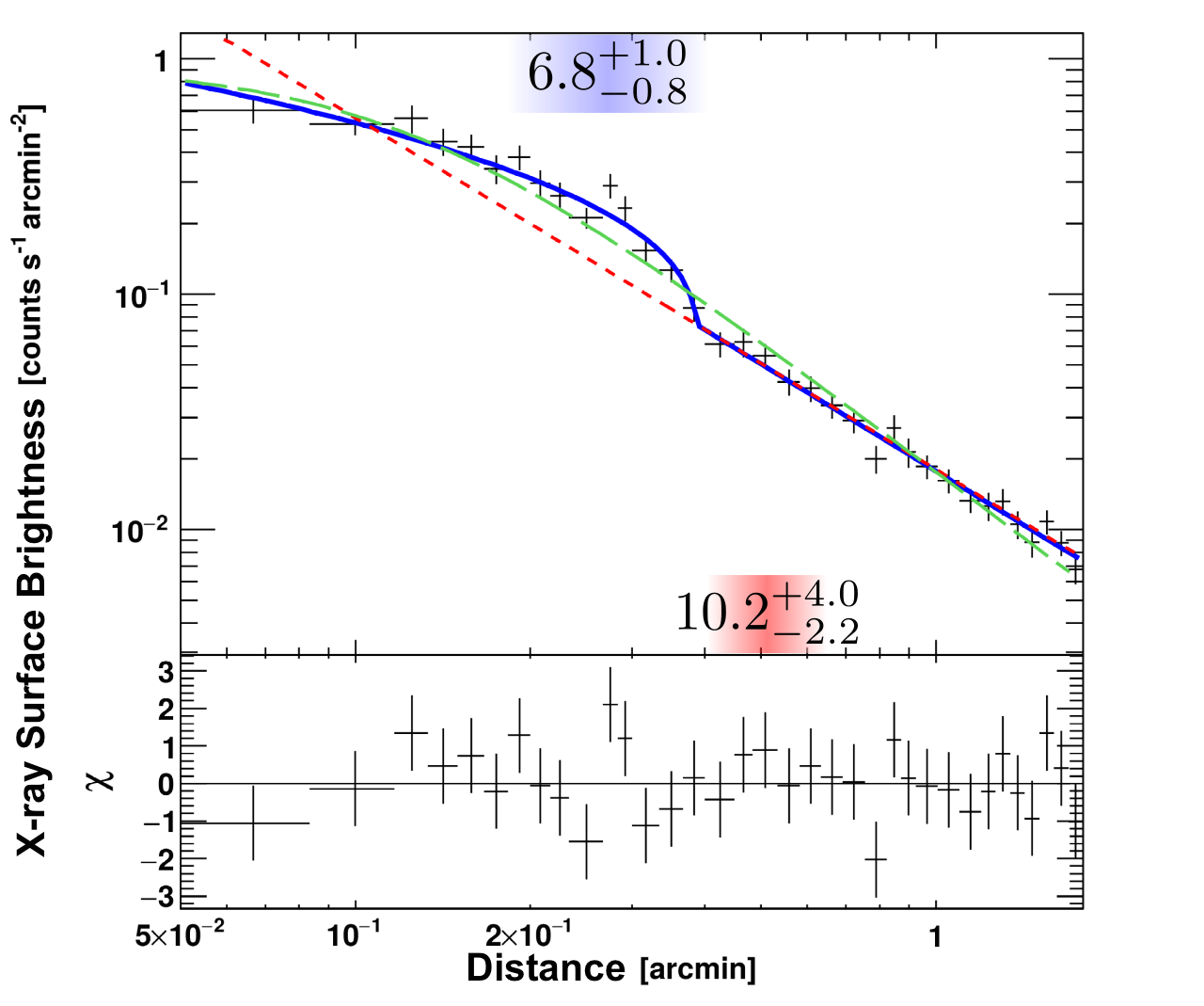}
        \includegraphics[width=0.5\textwidth]{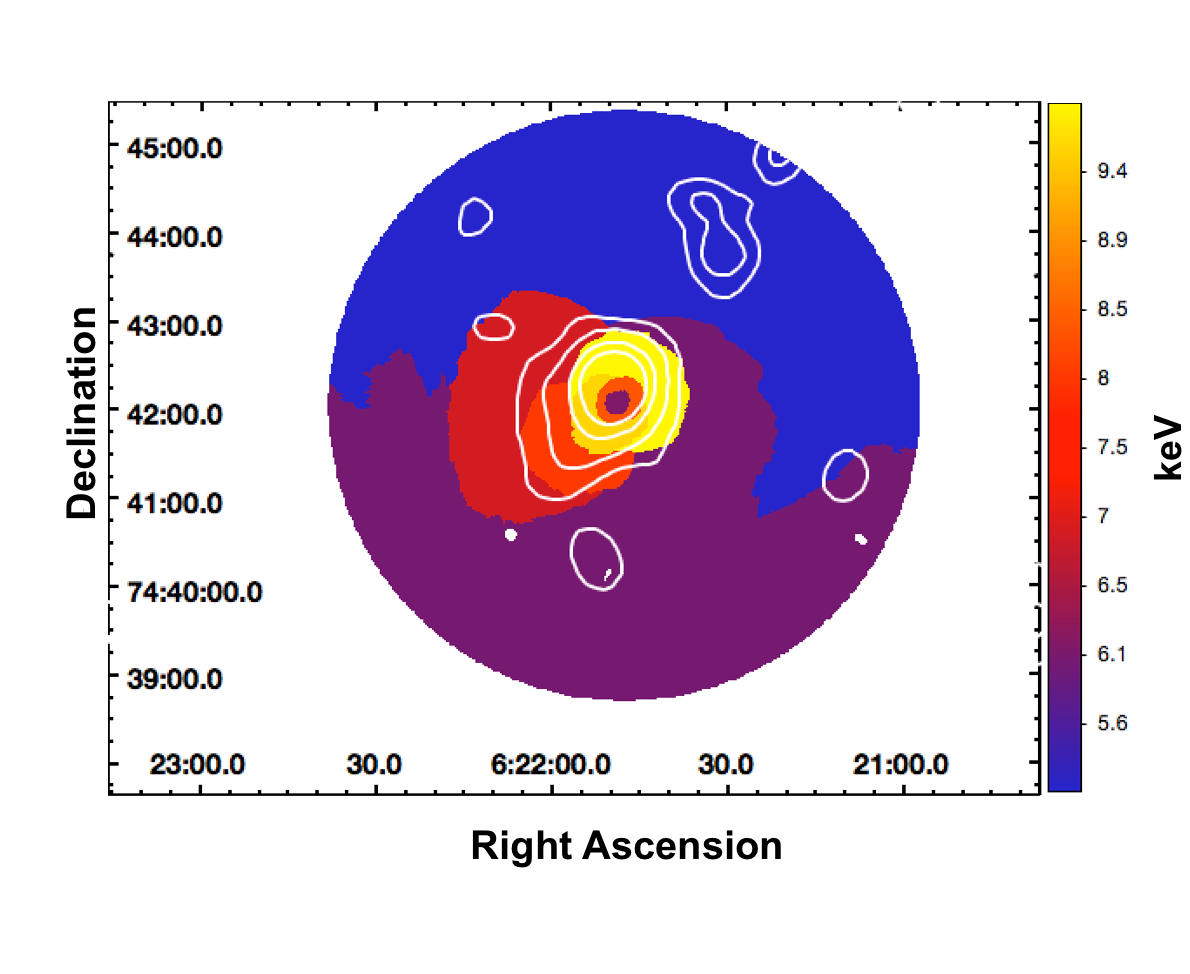}
    \caption{ 
    {\bf Top left panel:} Chandra X-ray image smoothed on a scale of 6$''$ with the overlay of the LOFAR contours and the sector used for the profile extraction. The inner grey arc indicates the position ($\sim 100$ kpc from the centre) of the cold front that we have discovered. Radio emission extends far beyond the cold front.  
    {\bf Top right panel:} X-ray surface brightness profile of the discontinuity detected in the Chandra image. The data were rebinned to reach a minimum signal-to-noise ratio of 7, and fitted with three models: broken-power law in solid blue ($\chi^2/dof$ = 26.5/29), power-law in dashed red ($\chi^2/dof$ = 152.1/31), and beta-model in dashed green ($\chi^2/dof$ =  67.8/31). The residuals at the bottom of the plot refer to the broken-power law model. The two colored boxes indicate the temperature in keV in the upstream and downstream regions, and their sizes indicate the radial extension of the spectral region. 
    {\bf Bottom panel:} Projected temperature map of the cluster with the contour levels at $(1, 2, 4, 8)\, \times \, 3\sigma$ where $\sigma$ = 500 $\mu$Jy/beam of the 144 MHz LOFAR image with a beam of 35$'' \times$ 35$''$ overlaid. The removed background sources (in white) are indicated. The presence of a cool core can be clearly seen. }
    \label{cold}
\end{figure*}

We reprocessed Chandra observations of the cluster in the 0.5-2.0 keV band to search for possible surface brightness and temperature jumps. We obtained a projected temperature map of the ICM using CONTBIN v1.4 \citep{Sanders2006} shown in the bottom panel of Fig. \ref{cold}. Given the short exposure of the Chandra observations, we required ~900 background-subtracted counts per bin. This resulted in 8 spectral regions that were then extracted and fitted. The temperature map clearly shows the presence of the cool core, and indicates a temperature jump from the central regions outwards in the NW direction. We then obtained the surface brightness profile and found a discontinuity towards the NW direction. The profile extraction and fitting were performed with PROFFIT \citep{EMP2011} on the exposure-corrected image of the cluster. The spectral profile was extracted in the sector indicated in grey in the top left panel of Fig.\ref{cold} and fitted with three models: broken-power law, power-law, and beta-model. The profile is best described by a broken power-law with compression factor $C = 1.7 \pm 0.1$. The spectral analysis provides temperatures of $6.8_{-0.8}^{+1.0}$ keV and $10.2_{-2.2}^{+4.0}$ keV in the upstream and downstream regions, respectively. In general, both shocks and cold fronts are sharp surface brightness discontinuities \citep{MV2007}, however shocks mark pressure discontinuities where the gas is heated in the downstream region, with higher temperature values with respect to the upstream region, whilst cold fronts do not show a pressure jump and the downstream temperature is lower than the upstream temperature. We have combined the temperature jump and compression ratio to estimate the pressure ratio across the edge and checked that the pressure is continuous across the front, as expected in the case of a cold front.\\

Given the discontinuity found both in the emissivity and temperature profiles, the morphology of the X-ray emission (elongated in the NW-SE direction) and the presence of a cool core, we argue that the cold front scenario is the most likely interpretation of the discontinuity.

\begin{figure}
	\includegraphics[width=0.5\textwidth]{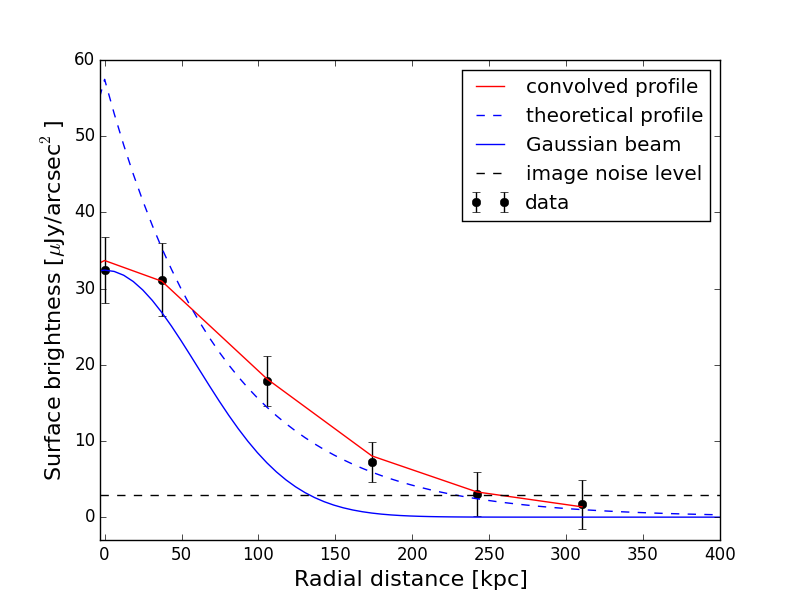}
 	\caption{The azimuthally-averaged brightness profiles of the radio emission in PSZ139. The best-fit line is indicated in red, and the 3$\sigma$ noise level of the radio image is indicated with a horizontal dashed-dotted black line. The best-fit values are $r_e = 94 \pm 10$ kpc and $I_0$ = $ 48 \pm 2 \, \mu$Jy/arcsec$^2$.}
    \label{profile}
\end{figure}

\section{Discussion and summary}

The information from the literature and the X-ray analysis that we have performed indicate that PSZ139 shows typical features of a non-merging cluster, i.e. a cool core with low central entropy. The X-ray morphology of the cluster on larger scales, though, is not spherically symmetric, suggesting the occurrence of a merger that has left a clear imprint on the gas distribution towards the SE of the cluster core. The morphological indicators $w$ and $c$ also suggest a cluster with a dense core and an elongated shape. Moreover, the radio mini halo discovered by \citet{Giaci2017} suggests that the core has not been disrupted. These features suggest a scenario where the cluster is undergoing a merger that is not powerful enough to disrupt the cool core. In fact, simulations have shown that it is quite hard for mergers to destroy cool cores in galaxy clusters \citep{Bu2008}.\\

Based on the correlation in \citet{Cassa2013} for giant halos, which compares cluster mass, halo radio power, and halo size, a cluster with the mass of PSZ139 is expected to host a giant halo with a power of $ P_{\rm tot,1.4} \sim 2 \times 10^{24}$ W Hz$^{-1}$ and a total size of $D_{\rm radio} \sim 950$ kpc.
The total estimated radio power at 144 MHz is $P_{\rm tot, 144} = (7.2 \pm 1.1) \times 10^{24}$ W/Hz, which corresponds to $P_{\rm tot, 1.4} = (3.7 \pm 0.5) \times 10^{23}$ W/Hz at 1.4 GHz, assuming a conservative spectral index of $\alpha = -1.3$ for the whole source. Even considering a large scatter around this correlation, the diffuse emission of PSZ139 is more than an order of magnitude underluminous and almost a factor of 2 smaller in size.\\

We argue that PSZ139 is the first example of a cluster that hosts ultra-steep-spectrum radio halo emission outside of its cool core. This radio emission is detected only at low frequencies and extends out to smaller radii than in typical giant radio halos. The radio analysis suggests that the more compact emission, which is coincident with the cool core and has a spectrum typical of mini halos ($\alpha_{144}^{610} \sim -1.3$), is surrounded by ultra-steep diffuse emission ($\alpha_{144}^{610} < -1.7$) that is correlated with the X-ray morphology on larger scales\footnote{We note that ultra-steep spectrum emission can also originate from the old lobes of a radio galaxy but we consider such a scenario unlikely, since we observe a strong spatial correlation between the radio and X-ray emission, as is typically seen in radio halos.}. The X-ray analysis indicates the presence of a cool core with traces of dynamical activity, especially towards the SE, in line with the direction where most of the larger-scale diffuse emission is detected. The presence of a cold front suggests that the core is likely to be sloshing. This observational evidence motivates us to identify a new scenario that links cool-core clusters and particle re-acceleration on cluster scales. We argue that the radio emission found in PSZ139 is caused by a minor merger that has dissipated enough energy in the ICM to accelerate particles, but leaves the core intact, as indicated by the low central entropy and temperature. Through the same mechanism that generates giant radio halos in merging clusters, less energetic mergers are predicted to form halos with a steeper spectrum and lower power than more energetic (major) mergers (\citealp{Cassa2006}, \citealp{Bru2008}). Up until now, this population of minor-merger, cool-core clusters remains largely undetected, and PSZ139 may in fact be the first example.
The radio emission of PSZ139 suggests that the energy dissipated during a minor merger can drive turbulent motions outside the cluster core, but on scales smaller than that of giant halos, whilst still preserving the mini halo emission. We exclude the scenario where the source in PSZ139 is a transition object: either that it is a giant halo decaying into a mini halo, or a mini halo growing into a giant halo. For a giant radio halo to fade from the edges and shrink to its present size, the radiative losses, and hence the magnetic field strength, would have to be stronger at the edges than in the core of the cluster. This is considered unlikely. In the other case, we can exclude that the relativistic plasma has been transported from the core out to larger scales since this would require unrealistically high transport coefficients.\\

The presence of diffuse emission on scales larger than the core has rarely been seen in non-merging clusters. Among the few known cases\footnote{Few cases of Mpc-scale halos have been found in massive clusters with no evidence of major mergers (CL1821+643, \citealp{Bona2014}; A2261, A2390, \citealp{Sommer2017}; A2142, \citealp{Ventu2017}). The former two sources substantially differ from PSZ139: CL1821+643 for being a flat-spectrum giant halo, and A2261 for not hosting a cool core. The latter two sources also differ from PSZ139, but deserve a more accurate comparison, and are therefore discussed in the main text.}, A2142 is the cluster that shares some similarities with PSZ139: it shows a two-component radio halo with flatter-spectrum emission in the core and a slightly steeper spectrum emission on Mpc scales \citep{Ventu2017}.
However, A2142 does not host a cool core, its halo has a size typical of giant radio halos, and the difference in the spectral index of the radio emission in the two components is only marginally significant. A case of a cool-core cluster hosting Mpc-scale emission is A2390 \citep{Sommer2017}, however the dynamic of the cluster is not clear, and the large errors on the spectral indices do not permit an assessment whether the emission on cluster scales has a steep spectrum.\\

PSZG139 is the first cool-core cluster to host steep-spectrum emission on larger scales. This example indicates that the connection between the evolution of radio emission on different scales and the dynamical status of the cluster is more complex than previously thought, and that particle acceleration mechanisms at different scales can be observed simultaneously in the same cluster.
As the energy dissipated by minor mergers is primarily observed at low radio frequencies, we expect that radio halos with steep spectra will be found in these types of clusters by forthcoming high-sensitivity, low-frequency radio observations.

\section*{Acknowledgements}
LOFAR, the Low Frequency Array designed and constructed by ASTRON, has facilities owned by various parties (each with their own funding sources), and that are collectively operated by the International LOFAR Telescope (ILT) foundation under a joint scientific policy. The LOFAR software and dedicated reduction packages on https://github.com/apmechev/GRID\_LRT were deployed on the e-infrastructure by the LOFAR e-infragroup, consisting of J.B.R.O. (ASTRON \& Leiden Observatory), A.P.M. (Leiden Observatory) and T.S. (Leiden Observatory) with support from N. Danezi (SURFsara) and C. Schrijvers (SURFsara). This research had made use of the LOFAR Solution Tool (LoSoTo), developed by F.dG., and of the NASA/IPAC Extragalactic Database (NED), which is operated by the Jet Propulsion Laboratory, California Institute of Technology, under contract with the National Aeronautics and Space Administration.
We thank the staff of the GMRT that made the observation possible. GMRT is run by the National Centre for Radio Tata Institute of Fundamental Research.\\
A.B. acknowledges support from the ERC-Stg 714245 DRANOEL. H.R. and R.vW. acknowledge support from the ERC Advanced Investigator programme NewClusters321271.
R.vW. acknowledges support of the VIDI research programme with project number 639.042.729, which is financed by the Netherlands Organisation for Scientific Research (NWO).
Basic research in radio astronomy at the Naval Research Laboratory is supported by 6.1 Base funding. G.W. gratefully thanks the Leverhulme Trust for funding.




\bibliographystyle{mnras}
\bibliography{PSZ} 




\bsp	
\label{lastpage}
\end{document}